\documentclass[acmtog]{acmart}
\acmSubmissionID{428}

\usepackage{booktabs} 
\usepackage{multirow}
\usepackage{enumitem}
\usepackage{cleveref}
\usepackage{amsmath}

\PassOptionsToPackage{capitalize}{cleveref}
\usepackage[ruled]{algorithm2e} 

\SetAlFnt{\small}
\SetAlCapFnt{\small}
\SetAlCapNameFnt{\small}
\SetAlCapHSkip{0pt}

\setcopyright{cc}
\setcctype{by}
\acmJournal{TOG}
\acmYear{2026} 
\acmVolume{45} 
\acmNumber{4} 
\acmArticle{140}
\acmMonth{7} 
\acmDOI{10.1145/3811313}

\begin{document}

\title{Semantic-Structural Alignment for Generative Pictorial Charts}

\author{Zhida Sun}
\orcid{0000-0003-4689-986X}
\email{zhida.sun@szu.edu.cn}
\affiliation{
   \department{Visual Computing Research Center (VCC), College of Computer Science and Software Engineering (CSSE)}
  \institution{Shenzhen University}
  \country{China}}

\author{Yulin Zhang}
\orcid{0009-0004-2770-9586}
\affiliation{
 \department{VCC, CSSE}
  \institution{Shenzhen University}
  \country{China}}

\author{Zheng Gu}
\orcid{0000-0001-9914-3922}
\affiliation{
 \department{VCC, CSSE}
  \institution{Shenzhen University}
  \country{China}}

\author{Min Lu}
\orcid{0000-0002-8464-0990}
\affiliation{%
 \institution{Shenzhen University}
 \country{China}}

\author{Bongshin Lee}
\orcid{0000-0002-4217-627X}
\affiliation{
  \institution{Yonsei University}
  \country{South Korea}}

\author{Daniel Cohen-Or}
\orcid{0000-0001-6777-7445}
\affiliation{%
  \institution{Tel Aviv University}
 \country{Israel}}

\author{Hui Huang}
\authornote{Corresponding author: Hui Huang (hhzhiyan@gmail.com)}
\affiliation{
	\department{VCC, CSSE}
	\institution{Shenzhen University}
	\country{China}}

\renewcommand\shortauthors{Z. Sun, Y. Zhang, Z. Gu, M. Lu, B. Lee, D. Cohen-Or, and H. Huang}

\begin{abstract}
  Traditional statistical graphics are precise but often lack the visual appeal, memorability, and engagement of pictorial charts. We present a generative framework for the automated synthesis of pictorial charts that bridges the gap between semantic expression and structural faithfulness. Rather than treating charts merely as images to be stylized, we frame the problem as a dual-conditioned generation task guided by two parallel external control signals: a text prompt capturing the semantic context of the editing intent, and a context image providing the abstract statistical chart's global structure. To reinforce these controls within a Multi-Modal Diffusion Transformer, we introduce two complementary feature-level mechanisms: structural alignment to anchor spatial layouts to the input chart, and semantic alignment to transfer expressive textures from reference images. Generalizing across major visual channels (i.e., length, area, angle, and position) and diverse semantic domains, our method produces pictorial charts that are both artistically compelling and structurally consistent. Extensive quantitative evaluations and perceptual user studies demonstrate that our framework outperforms traditional controllable generation and image editing baselines, providing a foundation for high-fidelity, data-driven generative modeling in expressive visual storytelling.
  Project page: \url{https://ssalign.github.io/}.
\end{abstract}


\maketitle

\section{Introduction}

\begin{figure}[t]
  \centering    
  \includegraphics[width=\linewidth]{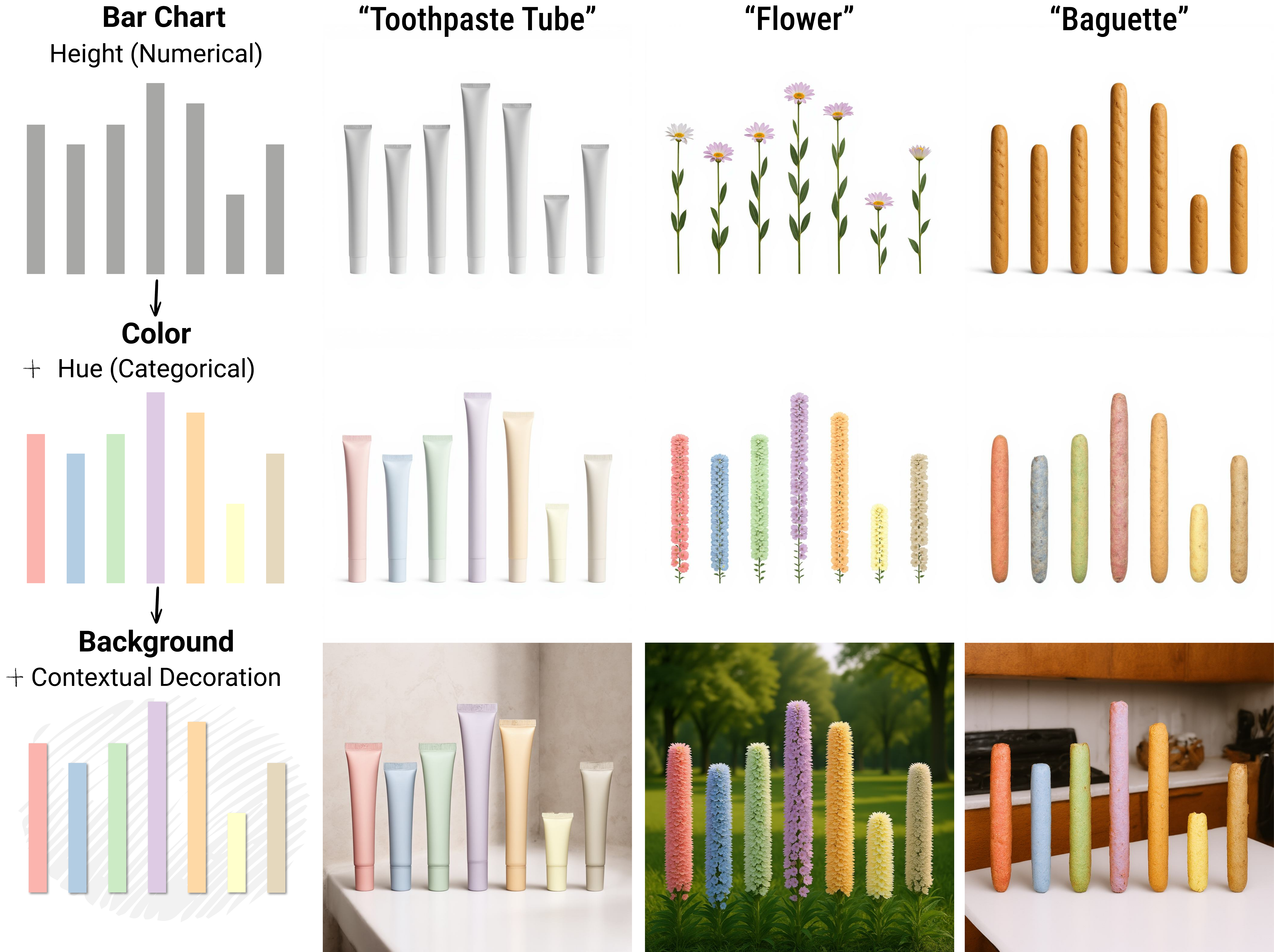}
    \caption{Semantically-Rich Pictorial Chart Generation.
    Our method transforms abstract statistical graphics into high-fidelity pictorial charts by replacing geometric primitives with text-guided semantic objects.
    It synthesizes instances structurally faithful to the original data encodings (e.g., proportional heights) and supports multi-channel visual encodings (e.g., categorical hues), which can be complemented with contextual backgrounds. This yields artistically compelling charts with strong structural consistency.
    }
  \label{fig:teaser}
\end{figure}

Pictorial charts transcend the generic geometric primitives of traditional statistical graphics by embedding quantitative information within recognizable visual referents, leveraging semantic grounding to bridge the gap between numerical data and human intuition~\cite{infomages}. 
By transforming abstract statistics into aesthetically resonant thematic imagery, these representations have been shown to enhance memorability and engagement~\cite{9903511}.
Consider the task of converting traditional bar charts into pictorial charts, where abstract rectangles are replaced by semantically meaningful objects (see Fig.~\ref{fig:teaser}).
In this setting, the generated pictorial elements are not merely decorative as their spatial extent serves as the primary carrier of numerical information, making precise geometric correspondence between the original bars and the generated objects a critical requirement. 
Therefore, our goal is to generate expressive imagery that maintains strong structural fidelity to the input data, ensuring that the alignment of generated objects preserves quantitative readability.

Despite their communicative efficacy, creating high-fidelity pictorial charts is traditionally limited to labor-intensive manual workflows~\cite{10.1145/3313831.3376172} or restricted by rigid, template-based systems~\cite{ChartGalaxy}.
Although rule-based tools and recent generative models have attempted to automate this process, they typically yield disjointed results that still necessitate extensive manual intervention to reconcile creative expression with statistical legibility~\cite{10.1109/TVCG.2023.3326913}.
This underscores the need for an automated generation process that synthesizes sophisticated pictorial charts without compromising either design nuance or data-encoding integrity.

Although the task of replacing abstract geometric primitives with semantic objects (see Fig.~\ref{fig:teaser}) appears straightforward, it exposes a limitation of current generative methods due to the lack of fine-grained, object-level control over spatial extent.
In diffusion architectures, object size is an entangled, emergent property rather than a manipulable variable, making off-the-shelf solutions fail to satisfy robust structural requirements.
Dense spatial guidance methods like ControlNet~\cite{controlnet} enforce rigid one-to-one correspondence, preventing the local deformations necessary to adapt organic shapes to data boundaries.  
Conversely, stochastic editing approaches such as SDEdit~\cite{sdedit} suffer from stochastic structural drift, compromising the structural fidelity required for visualization. 
Thus, achieving precise, multi-instance structural control without sacrificing semantic coherence remains an open challenge in fine-grained image synthesis.

In this work, we present a generative method designed to unify semantic appearance transfer with structural constraints on graphical charts.
Our approach utilizes a dual-conditioned generation framework that steers a diffusion model via two parallel signals.
Specifically, semantic control maps textual prompts into expressive visual referents, while structural control enforces geometric faithfulness by translating visual channels into localized geometric transformations.
At the core of our method is an explicit semantic and structural disentanglement implemented through two feature-level alignment modules, Structural DIFT and Semantic DIFT, which jointly enable expressive pictorial synthesis while preserving structural fidelity.
Finally, this automated semantic-structural alignment logic evaluates and optimizes the topological compatibility between the chosen referent and the chart, yielding pictorial charts that are both artistically compelling and structurally consistent.

To clearly illustrate our method, we focus our primary demonstration on standard bar charts while further validating its extensibility. 
Organized around four fundamental visual encoding channels (e.g., length, area, angle, and position), our framework generalizes to unseen chart forms sharing the same structural encoding logic. 
However, to ensure rigorous structural consistency, we limit our scope to the one-to-one semantic transformation of primary geometric primitives, excluding highly nested or multi-view visualizations.
Accordingly, our focus remains on transforming the data-encoding chart elements themselves, deferring the distinct challenge of cohesive, prompt-driven background generation to future investigations.

\section{Related Work}

\paragraph{Pictorial Charts Creation}
The creation of pictorial charts, which embed quantitative data into semantically rich visual representations, has evolved through manual authoring, example-based retrieval, and recent generative approaches.
Early research prioritized \textit{interactive authoring}, where systems such as Data Illustrator~\cite{10.1145/3173574.3173697}, Data-Driven Guides~\cite{Data-DrivenGuides}, and DataInk~\cite{10.1145/3173574.3173797} enabled the freehand drafting of glyphs and precise adjustment of graphical attributes.
However, while these tools facilitate highly customized output, they require design expertise and impose a high cognitive load, limiting scalability.
To alleviate this burden, subsequent work explored example-driven synthesis, exemplified by methods like DataQuilt~\cite{10.1145/3313831.3376172}, Infomages~\cite{infomages}, DataWink~\cite{DataWink}, and deep-learning-based template extraction~\cite{InfographicDesign}, which automate creation via the retrieval and substitution of imagery assets or thematic backgrounds. 
Yet, these approaches remain fundamentally constrained by the availability of curated libraries and often suffer from semantic-geometric mismatches, where retrieved assets cannot adapt to dynamic data distributions without distortion.
Most recently, generative approaches utilizing latent diffusion models, such as Vistylist~\cite{9903511}, viz2viz~\cite{wu2023viz2vizpromptdrivenstylizedvisualization}, and ChartSpark~\cite{10.1109/TVCG.2023.3326913}, have demonstrated potential for aesthetic style transfer, stylized visualization generation, and semantic asset generation.
However, these methods only incorporate chart structure as a hard constraint and require complex technical combinations to enrich semantics, while lacking flexible generation that maintains structural consistency.
In contrast, our dual-conditioned mechanism automatically resolves this trade-off by enforcing global topological constraints and semantic alignment, ensuring that the generated objects serve as functional and accurate data encoders.

\paragraph{Controllable Image Generation}
Achieving precise control over the synthesis process remains a challenge in generative modeling.
Existing methods generally achieve this either by optimizing additional model weights (training-based) or by modulating the inference process (training-free).
\textit{Training-based approaches}, such as ControlNet~\cite{controlnet}, provide robust guidance but
impose rigid, one-to-one structural constraints. 
Consequently, they prohibit the local geometric deformations required to transform abstract chart primitives into organic semantic objects.
Conversely, \textit{training-free methods} modulate attention maps directly during inference.
Approaches like Cross-Image Attention (CIA)~\cite{Cross-Image-Attention}, Ctrl-X~\cite{Ctrl-X}, and Attention Distillation~\cite{AttentionDistillation} guide structure and appearance by manipulating, optimizing, or using attention features from reference images. 
However, when the semantically rich target domain diverges significantly from the abstract structure-only source domain, these methods struggle to maintain structural fidelity.
To overcome this limitation, we advance the attention-based paradigm by integrating diffusion feature-level correspondence into the generative process. 
This approach achieves a flexible balance between global structural consistency and local semantic adaptation.

\paragraph{Structure-Guided Generative Appearance Transfer}
In contrast to unconditional generation, structure-guided editing aims to decouple a target's structure (e.g., spatial layout, geometry) from its appearance (e.g., style, texture, identity) to transfer new visual semantics. We categorize existing approaches into noise-based guidance, attention-based manipulation, and context-based editing.
Early \textit{noise-based techniques} (e.g., SDEdit~\cite{sdedit}) utilize stochastic noise injection to maintain global structure, while Rectified Flow methods like RF-Edit~\cite{RF-Edit} improve inversion fidelity by treating the generative process as a deterministic Ordinary Differential Equation (ODE). 
However, these global inversion techniques often struggle to cleanly decouple geometry from texture.
\textit{Attention-based methods} such as MasaCtrl~\cite{MasaCtrl}, Cora~\cite{Cora}, and Stable Flow~\cite{StableFlow} manipulate self-attention layers or specific DiT blocks to retain spatial layouts while swapping appearance, though they suffer from limited editability when the structural domain shifts significantly.

Most recently, \textit{context-based approaches}~\cite{ICLoRA, EditTransfer, ICEdit} have leveraged the architectural strengths of Diffusion Transformers (DiT). 
These works demonstrate that DiT models possess an inherent capability to perceive global image context, allowing source images to serve as flexible contextual conditions. 
This is critical, as it avoids constraining the editing results to the rigid pixel distribution of the input, facilitating versatile transformations. 
In this work, we build upon this context-aware paradigm to address the rigorous structural fidelity required for pictorial chart generation.

\section{Preliminaries}

\paragraph{MultiModal Diffusion Transformers}
Rectified-flow Diffusion Transformers (DiTs) model the generative process as an ODE transport along a linear trajectory between data and noise, defined by:
\begin{equation}
{z}_t = (1-t) {x}_0 + t \boldsymbol{\epsilon}, \quad \boldsymbol{\epsilon}\sim\mathcal{N}({0},{I}),
\end{equation}
where $x_0$ denotes the data sample, $z_t$ is the latent variable at timestep $t\in[0, 1]$, $\boldsymbol{\epsilon}$ is Gaussian noise, and $I$ is the identity matrix.
The model is optimized via conditional flow matching, which regresses the target vector field driving this interpolation.
To condition the generation process on textual inputs, the MultiModal Diffusion Transformer (MM-DiT) tokenizes the noised latent image into patch tokens and the text prompt into a sequence of text tokens. 
These two token streams are concatenated into a joint token sequence ${H}\in\mathbb{R}^{n\times d}$. The multi-modal self-attention is then computed as:
\begin{equation}
    \mathrm{Attn}({Q,K,V})=\mathrm{Softmax}\!\left(\frac{{Q}{K}^{\top}}{\sqrt{d_h}}\right){V},{Q}={H}{W}_Q,{K}={H}{W}_K,{V}={H}{W}_V,
\end{equation}
where $d_h$ denotes the per-head channel dimension, $H$ denotes the joint sequence of image and text tokens, $n$ is the sequence length, $d$ is the token feature dimension, $Q$, $K$, and $V$ denote the query, key, and value matrices, respectively, and $W_Q$, $W_K$, and $W_V$ are the learnable projection matrices mapping the joint sequence $H$ into the query, key, and value matrices, respectively.

\begin{figure}[t]
  \centering    
  \includegraphics[width=0.92\linewidth]{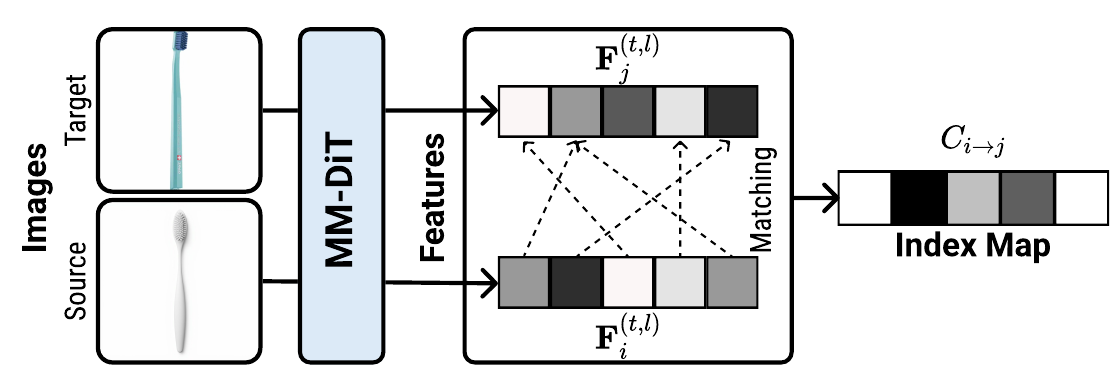}
    \caption{Token-Level Correspondence via DIFT. Given two images, diffusion features are extracted from MM-DiT at a specific denoising timestep $t$ and layer $l$, where each feature corresponds to a visual token on the latent grid. Token-level correspondence is then established by nearest-neighbor matching in the feature space using cosine similarity, resulting in a dense index map that represents cross-image correspondence. 
    }
  \label{fig:dift}
\end{figure}

\paragraph{Diffusion Features}
Pretrained diffusion models internalize rich semantics that enable robust, off-the-shelf correspondence estimation. During the denoising process, intermediate Diffusion Features (DIFT)~\cite{DIFT} inherently encode both local appearance and global structure, facilitating consistent matching across diverse images.
Specifically, let ${F}_i^{(t,l)}\in\mathbb{R}^{N_i\times C}$ and ${F}_j^{(t,l)}\in\mathbb{R}^{N_j\times C}$ denote two diffusion features extracted from layer $l$ at timestep $t$, where each row corresponds to a visual token (i.e., a latent-grid location after patchification).
We define the token-level correspondence map $C_{i\to j}$ via nearest-neighbor retrieval using cosine similarity:
\begin{equation}
C_{i\to j}(v_i)
=\arg\max_{v_j\in\mathcal{V}_j}\ \operatorname{sim}\!\left({F}_i^{(t,l)}(v_i),\,{F}_j^{(t,l)}(v_j)\right),
\end{equation}
where $v_i\in\mathcal{V}_i$ and $v_j\in\mathcal{V}_j$ index the visual tokens of images ${I}_i$ and ${I}_j$, and $\operatorname{sim}$ denotes the cosine similarity.
For brevity, we subsequently denote ${F}_*^{(t,l)}$ simply as ${F}_*$.
This formulation extracts a dense feature map where each spatial location serves as a descriptor for inter-image correspondence estimation (see Fig.~\ref{fig:dift}), providing the foundation for our dual-conditioned structural and semantic alignment.

\section{Methods}

\begin{figure}[t]
  \centering    
  \includegraphics[width=\linewidth]{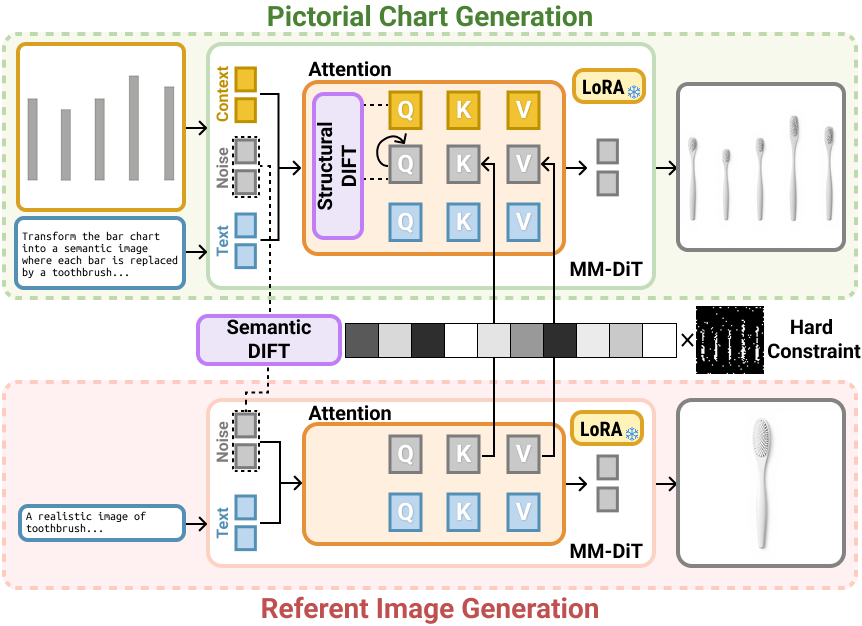}
    \caption{Method overview. 
    Given a source chart image and a textual prompt, we provide them as the contextual image condition and textual condition, respectively. To enforce strong structural consistency and enable expressive semantic synthesis, we introduce \emph{Structural DIFT} and \emph{Semantic DIFT}. 
    These operations are performed within the self-attention layers of the single-stream blocks in the MM-DiT.
    }
  \label{fig:case}
\end{figure}

Given an input statistical chart and a textual editing prompt, our goal is to synthesize a high-fidelity pictorial chart that faithfully encodes the original quantitative data into semantically meaningful visual referents (Fig.~\ref{fig:case}).
The pipeline begins with data preparation and model adaptation, where we fine-tune an MM-DiT via Low-Rank Adaptation (LoRA) on a combination of curated and augmented datasets to align abstract primitives with naturalistic objects.
At inference, we introduce a dual-branch mechanism inspired by DIFT, consisting of a Structural DIFT module that enforces robust structural alignment with the input layout, and a complementary Semantic DIFT module that guides semantic synthesis through dense feature correspondence.

\subsection{Structure-Aware Semantic Transformation}

\begin{figure}[t]
    \centering    
    \includegraphics[width=\linewidth]{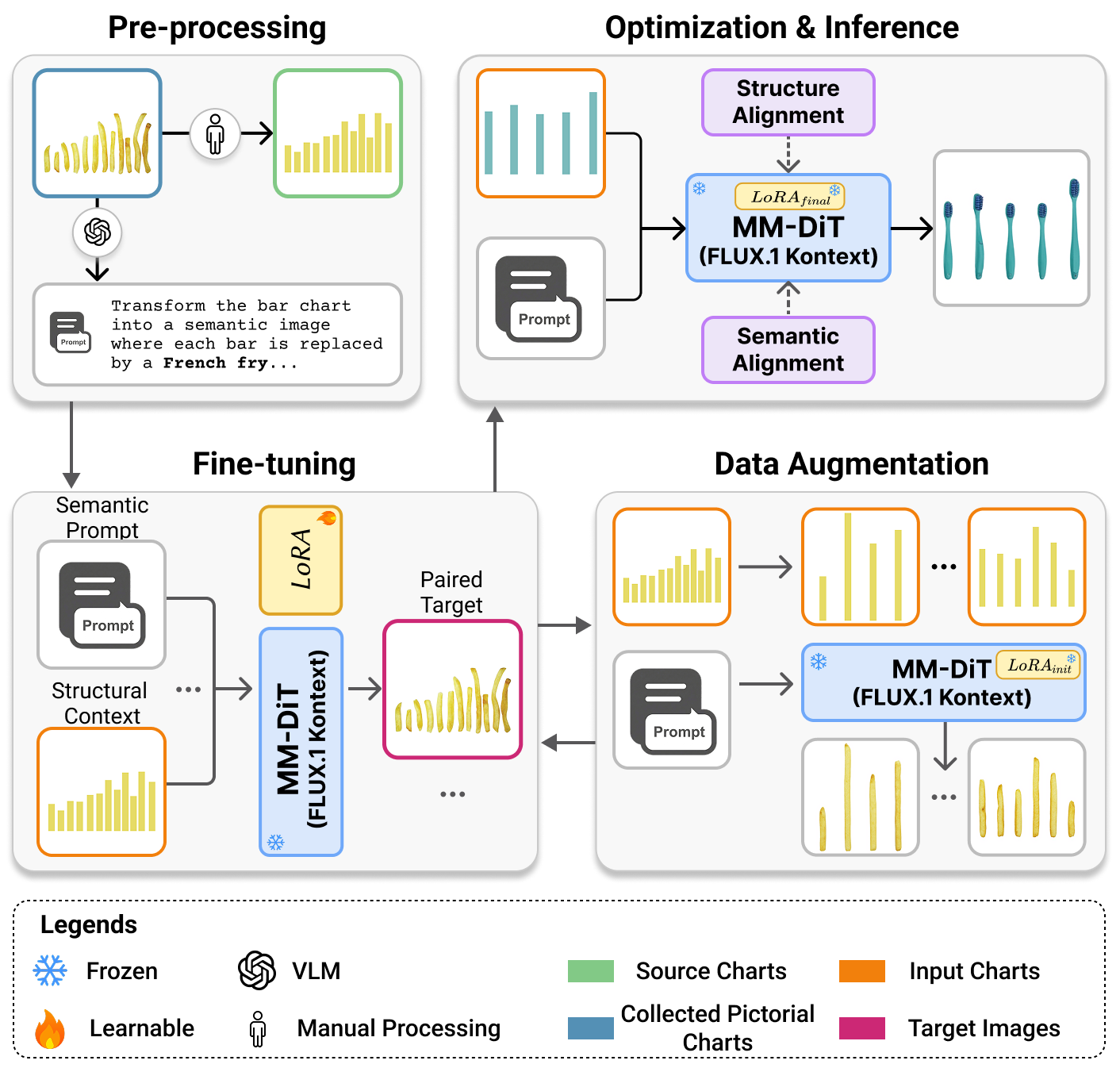}
    \caption{
    Training pipeline. 
    Overview of our progressive data curation and fine-tuning strategy:
    (a) Data Preprocessing: We pair manually collected pictorial charts with reverse-engineered geometric source charts, generating corresponding text prompts via a Vision-Language Model (VLM).
    (b) Fine-Tuning \& Data Augmentation: A baseline LoRA is fine-tuned on this small seed dataset and subsequently utilized to autonomously synthesize a large-scale, structurally diverse augmented dataset.
    (c) Optimization \& Inference: The augmented data spanning multiple visual encoding channels is aggregated for a final, comprehensive LoRA fine-tuning, equipping the model with a robust generative prior for expressive semantic synthesis.
    }
  \label{fig:pipeline}
\end{figure}

Since statistical graphics composed of geometric primitives exhibit limited texture and weak semantic content, standard generative models struggle to bridge the domain gap.
They often fail to synthesize detailed textures onto sparse primitives or, conversely, destroy the strong structural fidelity required for statistical charts.
To resolve this, we fine-tune a DiT to enable structure-aware semantic transformation, as illustrated in Fig.~\ref{fig:pipeline}.
We build upon FLUX.1 Kontext, a model pre-trained on extensive editing pairs that naturally supports contextual image conditioning. 
However, a critical bottleneck is the lack of paired training data, as existing collections like PiCCL~\cite{PiCCL} provide only the final pictorial charts ($I_{tgt}$) without their corresponding source charts ($I_c$). To address this, we curated 120 high-quality pictorial charts from public datasets and manually reverse-engineered the underlying source chart for each, reconstructing the precise spatial structure to form valid pairs. 
For each sample, we leveraged a vision-language model to extract semantic attributes and generate candidate editing prompts, which were subsequently verified by human annotators.
This process yielded an initial high-quality dataset of 583 paired samples in the form $(I_{c}, \text{prompt}) \rightarrow I_{tgt}$.
Crucially, to ensure their spatial structure remains a faithful carrier of the numerical data, the charts in our dataset encode values through the continuous dimensions of the pictorial elements (e.g., height, length, area, and position) rather than through discrete element counts.

To further enhance the model's sensitivity to structural variations and mitigate the scarcity of training data, we devised a structure-driven augmentation strategy. 
We partitioned the dataset into four distinct categories based on the primary visual encoding channel: length, area, angle, and position. 
For each category, we trained a provisional LoRA model using the manually curated pairs. 
We then procedurally perturbed the geometric parameters of the source charts to create structurally diverse inputs and utilized the category-specific LoRA to synthesize corresponding target images. 
This bootstrapping process generated a large set of synthetic pairs that are semantically consistent with the original style but structurally varied. 
Finally, we aggregated these augmented samples with the original dataset to train the final LoRA, ensuring the model generalizes robustly across diverse chart topologies.

\subsection{Structure Alignment}

\begin{figure}[t]
  \centering    
  \includegraphics[width=0.9\linewidth]{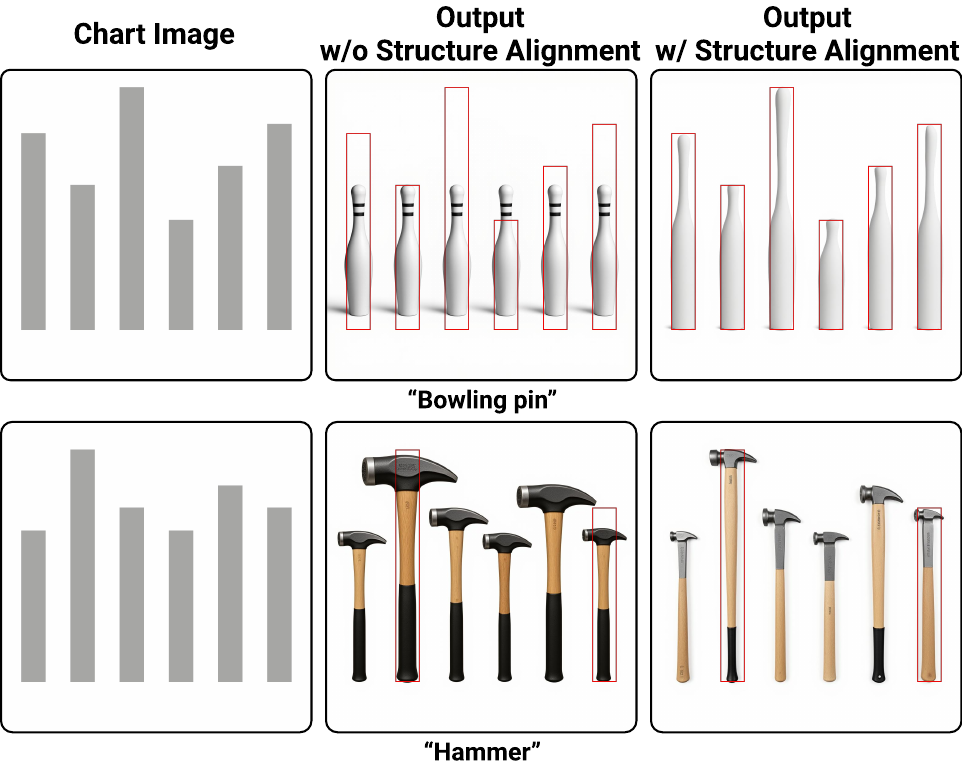}
    \caption{Structure Alignment. 
    While the fine-tuned MM-DiT provides a strong baseline, structural deviations from the source chart may still occur. By incorporating our proposed structural alignment strategy, the target image is explicitly aligned with the precise spatial structure of the chart, ensuring strong structural fidelity.
    }
  \label{fig:structurealign}
\end{figure}

\begin{figure}[b]
  \centering    
  \includegraphics[width=\linewidth]{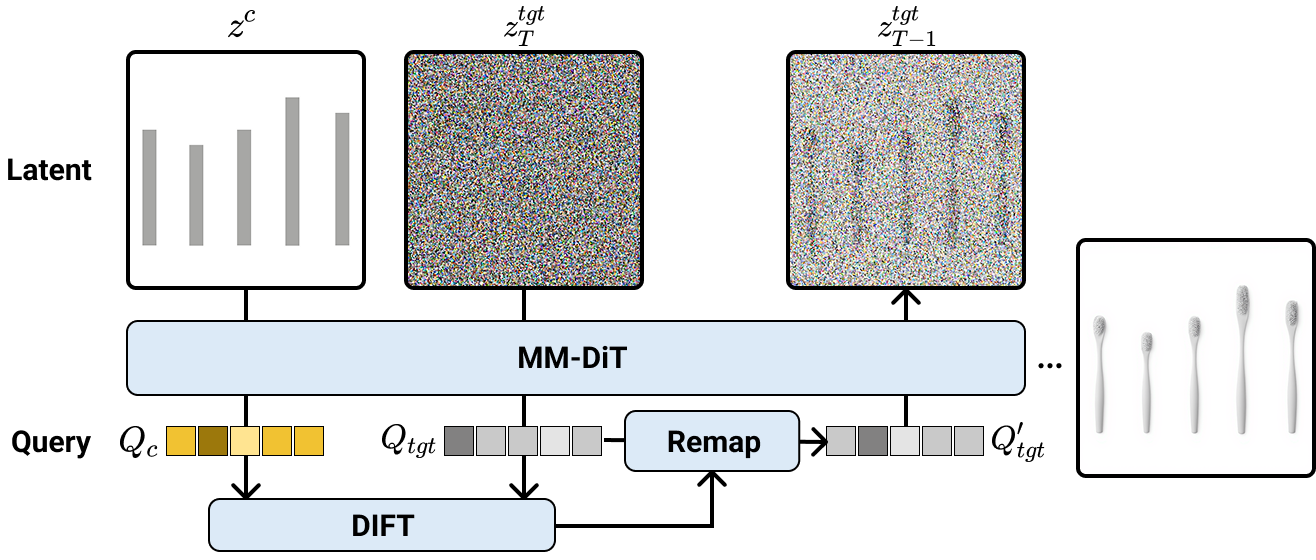}
    \caption{Structural DIFT. During the early stages of denoising, we compute the dense correspondence between the chart image queries $Q_c$ and the target image queries $Q_{tgt}$ using diffusion features. Based on these established matching relationships, we remap the target queries $Q_{tgt}$ to enforce strict structural alignment with the source chart.
    }
  \label{fig:structurealignarch}
\end{figure}

Despite the LoRA adaptation, the generated output may still exhibit structural deviations due to the limited scale of the training data and inevitable noise in the preprocessing pipeline.
To enforce strong structural fidelity to the underlying data encoding (Fig.~\ref{fig:structurealign}), we propose \textit{Structural DIFT}, a mechanism that explicitly aligns the spatial layout of the generated target $I_{tgt}$ with the source chart $I_c$ during inference (Fig.~\ref{fig:structurealignarch}). 
Our approach leverages two key insights: first, manipulating Query ($Q$) projections in self-attention layers effectively governs global image structure; and second, internal DIFTs provide a robust metric for establishing semantic-structural correspondence.

Specifically, Structural DIFT enforces alignment by rectifying the target queries $Q_{tgt}$ to match the spatial configuration of the chart queries $Q_c$. We define the token-level index map as:
\begin{equation}
C_{c\to tgt}(v_{c})=\arg\max_{v_{tgt}\in\mathcal{V}_{tgt}}\ \operatorname{sim}\!\left({Q}_{c}(v_{c}),\,{Q}_{tgt}(v_{tgt})\right),
\label{eq:DIFTstructure}
\end{equation}
where $v_{tgt} \in \mathcal{V}_{tgt}$ and $v_c \in \mathcal{V}_c$ index the visual tokens of the target image ${I}_{tgt}$ and the chart image ${I}_c$, respectively. 
To achieve this, we inject the source chart image $I_c$ as a context input within the visual stream.
Then, based on the index map $C_{c\to tgt}(v_{c})$, we spatially remap the target image queries ${Q}_{tgt}$ to achieve strict structural alignment between the target image and the chart (Fig.~\ref{fig:process}):
\begin{equation}
Q_{tgt}^{\prime} = Q_{tgt}\!\left(C_{c\to tgt}(v_{c})\right).
\label{eq:structure align}
\end{equation}
Crucially, this alignment is applied exclusively during the early denoising stages to establish the global layout without suppressing high-frequency semantic details. 
Furthermore, we restrict this operation to single-stream blocks on image-level tokens only, ensuring the alignment process does not interfere with the text-driven semantic control capabilities of the model.

\subsection{Semantic Alignment}

\begin{figure}[t]
  \centering    
  \includegraphics[width=\linewidth]{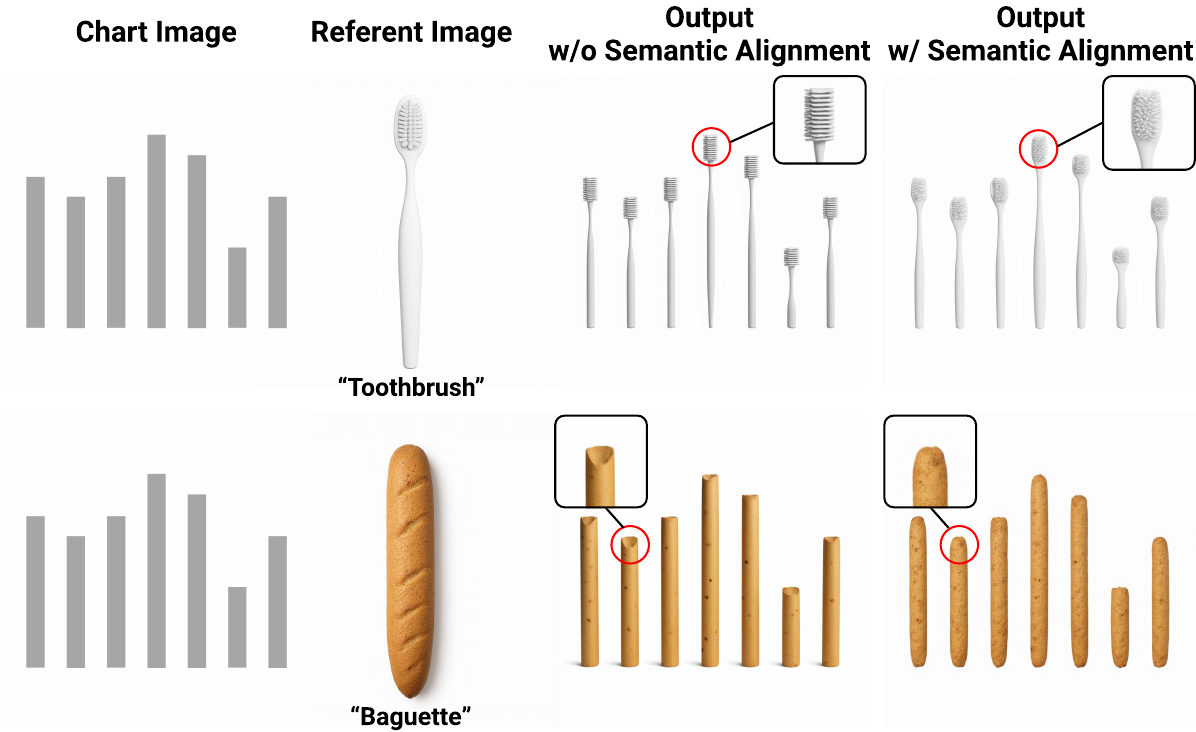}
    \caption{Semantic Alignment. Due to the limitations of few-shot fine-tuning, the model may occasionally produce unnatural artifacts or degraded visual referents. By incorporating a feature-level interpolation strategy using a reference image, we guide expressive semantic synthesis and effectively mitigate these limitations.
    }
  \label{fig:semanticalign}
\end{figure}

\begin{figure}[b]
  \centering    
  \includegraphics[width=\linewidth]{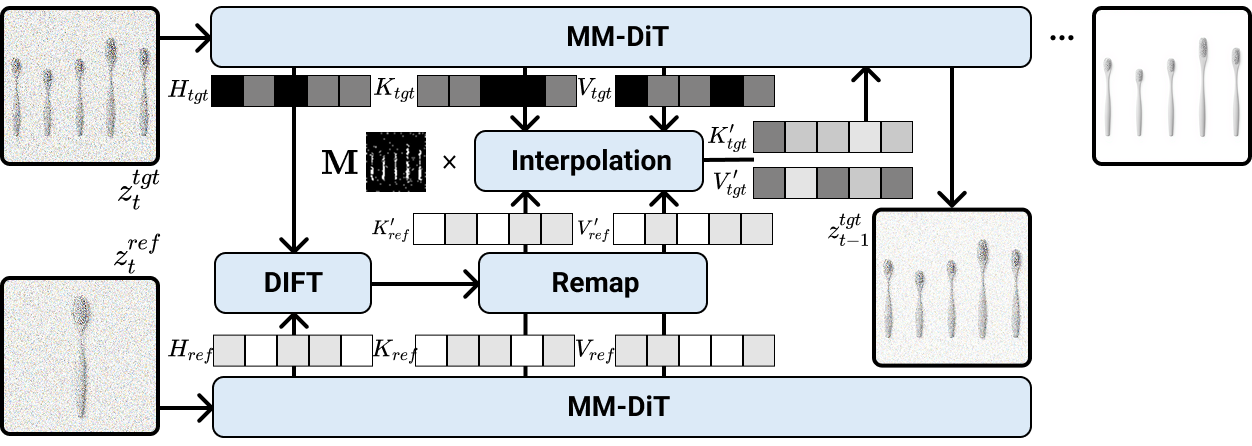}
    \caption{Semantic DIFT. We compute dense correspondences using diffusion features to spatially remap reference keys and values ($K_{\text{ref}}, V_{\text{ref}}$). These aligned features are then interpolated with the target ($K_{\text{tgt}}, V_{\text{tgt}}$) under a semantic mask $M$ to achieve expressive semantic alignment.}
  \label{fig:semanticdiftarch}
\end{figure}

Although the fine-tuned DiT model is capable of semantic transformation, few-shot adaptation can occasionally result in degraded object fidelity (Fig.~\ref{fig:semanticalign}).
Furthermore, the explicit injection of structural features from the previous stage may inadvertently introduce chart-like artifacts that interfere with the intended organic texture.
To resolve this and support user-customizable appearances, we introduce \textit{Semantic DIFT}, a mechanism that enforces semantic alignment by transferring appearance features from a reference image $I_{ref}$.

\begin{figure}[t]
  \centering    
  \includegraphics[width=1.\linewidth]{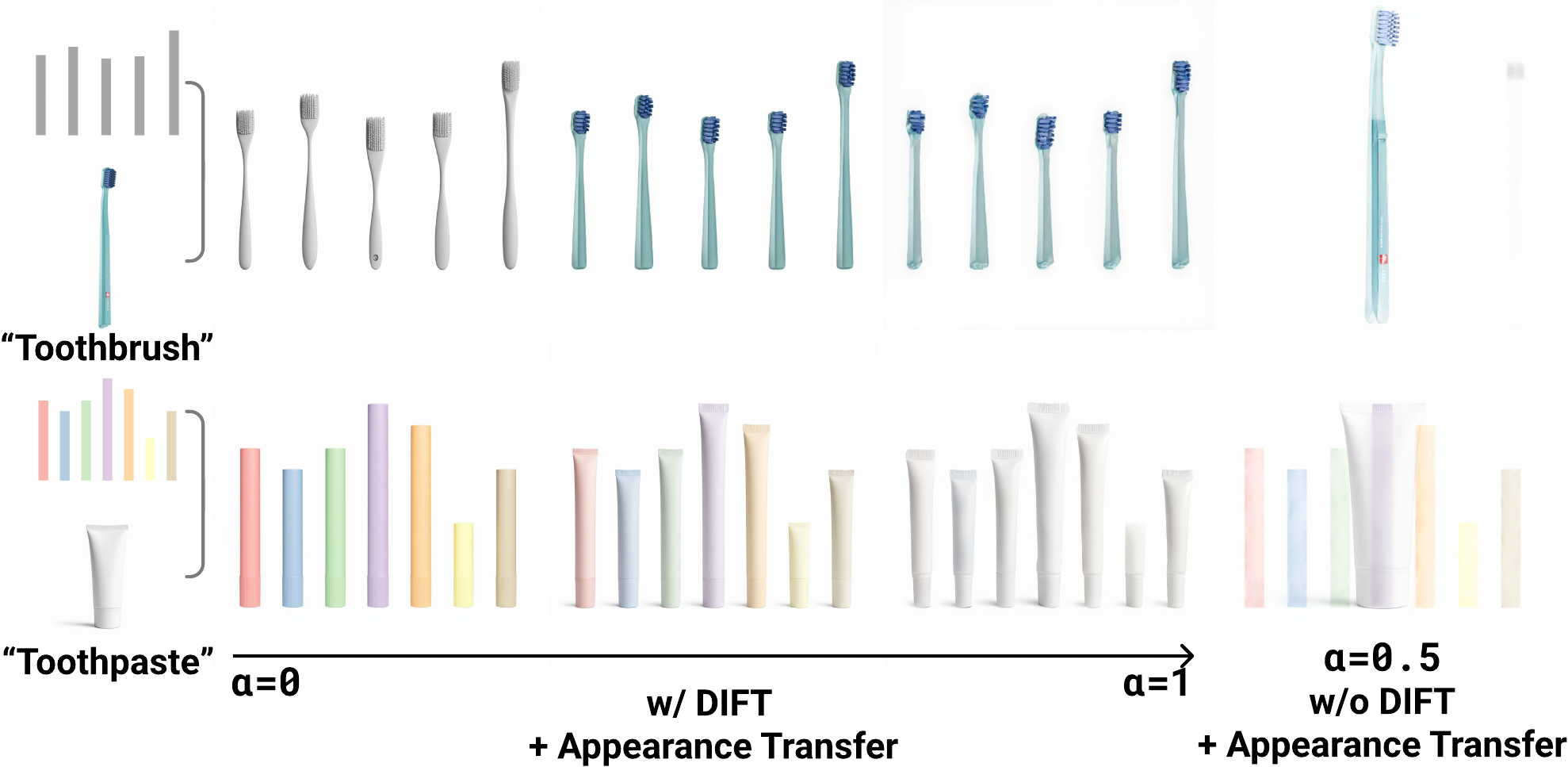}
    \caption{Effect of Semantic DIFT. DIFT-guided interpolation enables high-fidelity semantic synthesis. Crucially, it prevents the loss of critical data-encoding colors caused by full feature replacement ($\alpha = 1$) and mitigates structural artifacts arising from naive correspondence.
    Rows display results using inverted references (top) and text-generated references (bottom).}
  \label{fig:appdift}
\end{figure}

\begin{figure}[!h]
    \centering    
    \includegraphics[width=\linewidth]{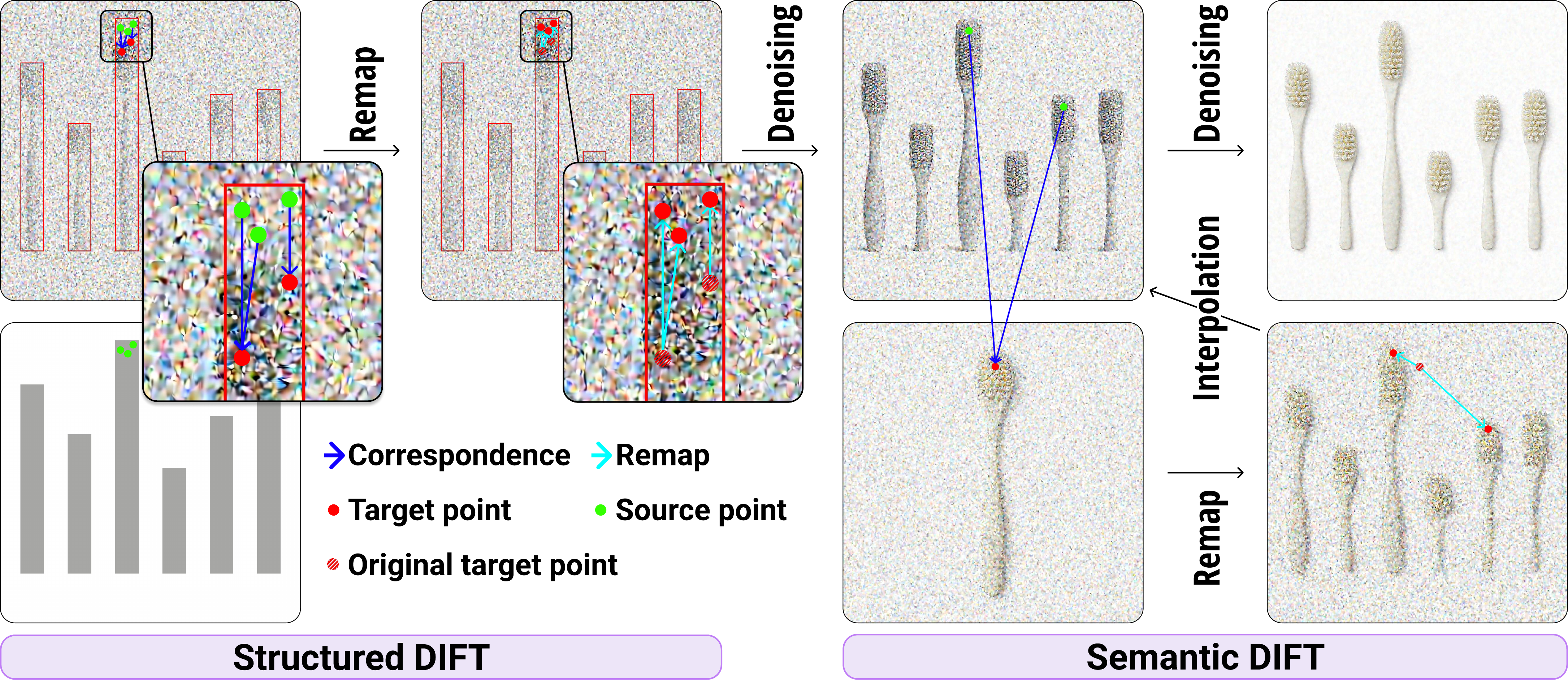}
    \caption{The DIFT Remapping Process. 
    (a) Structural DIFT: Dense correspondence ($C_{c\to tgt}$, blue arrows) is computed between the reference chart queries ($Q_{\text{c}}$, green points) and the target queries ($Q_{\text{tgt}}$, red points). 
    Based on this correspondence, the target features are spatially remapped (cyan arrows) to new positions, aligning their geometric layout with the structural arrangement of the reference chart.
    (b) Semantic DIFT: Using the established correspondence ($C_{tgt\to ref}$) between the target features ($H_{\text{tgt}}$, green points) and the reference features ($H_{\text{ref}}$, red points), the reference keys ($K_{\text{ref}}$) and values ($V_{\text{ref}}$) are spatially remapped. These are subsequently interpolated with the target keys ($K_{\text{tgt}}$) and values ($V_{\text{tgt}}$) to fuse the semantic attributes.
    }
  \label{fig:process}
\end{figure}

Prior appearance transfer methods developed for U-Net architectures~\cite{Cross-Image-Attention, Ctrl-X} typically rely on the direct substitution of Self-Attention Keys ($K$) and Values ($V$). 
However, within the DiT framework, we observe that relying solely on naive self-attention matching is insufficient due to inaccurate correspondence. 
To address this, Semantic DIFT computes dense semantic correspondences between the reference and the target images using diffusion features (Fig.~\ref{fig:semanticdiftarch}). 
We then spatially remap the keys $K_{ref}$ and values $V_{ref}$ of the reference image to align with the target's structure, ensuring that texture transfer respects the generated structural fidelity (Fig.~\ref{fig:process}). 
For brevity, we formulate this alignment for the keys $K$, with analogous operations applied to the values $V$:
\begin{equation}
C_{tgt\to ref}(v_{tgt})=\arg\max_{v_{ref}\in\mathcal{V}_{ref}}\ \operatorname{sim}\!\left({H}_{tgt}(v_{tgt}),\,{H}_{ref}(v_{ref})\right),
\label{eq:DIFTapp}
\end{equation}
\begin{equation}
K_{ref}^{\prime} = K_{ref}\!\left(C_{tgt \to ref}(v_{tgt})\right),
\label{eq:key align}
\end{equation}
where ${H}_{ref}$ and ${H}_{tgt}$ denote the features of the reference image $I_{ref}$ and the target image $I_{tgt}$, respectively, extracted prior to the self-attention layers, and $K_{ref}^{\prime}$ represents the remapped reference keys.
Instead of directly replacing the target features, which often introduces visual artifacts or overrides critical data-encoding colors, we employ Spherical Linear Interpolation (SLERP) to smoothly blend the remapped reference features with the target features (Fig.~\ref{fig:appdift}):
\begin{equation}
K_{tgt}^{\prime} =  \operatorname{SLERP}(K_{ref}^{\prime}, K_{tgt}; \alpha),
\label{eq:key interpolation}
\end{equation}
where $\alpha \in [0,1]$ controls the blending intensity.
To prevent the transferred appearance from spilling over the target spatial boundaries and compromising structural consistency, we constrain this transfer using a semantic mask $M$ derived from attention maps \cite{ConsistEdit, MasaCtrl}. 
The final update of the keys and values is formulated as:
\begin{equation}
K_{tgt}^{\prime}
=
K_{tgt}^{\prime}\odot {M}
+
K_{tgt}\odot\left(1 - {M}\right),
\quad V_{tgt}^{\prime}
=
V_{tgt}^{\prime}\odot {M}
+
V_{tgt}\odot\left(1 - {M}\right).
\end{equation}
Similar to structural alignment, this operation is applied exclusively within the single-stream blocks on visual tokens. 
Finally, the updated target image keys $K^{\prime}_{tgt}$ are concatenated with the original prompt keys $K_p$ and the chart image keys $K_c$:
\begin{equation}
K^{\prime} \leftarrow \mathrm{Concat}\!\left(K_p,\; K^{\prime}_{tgt},\; K_{c}\right),
\end{equation}
where $\mathrm{Concat}$ denotes the concatenation operation, and $K^{\prime}$ represents the complete sequence of keys. The same operation is applied to the values and queries, yielding the concatenated sequences $V^{\prime}$ and $Q^{\prime}$, respectively.
Consequently, the output of the self-attention mechanism at layer $l$ (from the single-stream block) is given by:
\begin{equation}
\mathrm{Attn}^{(l)}(Q^{\prime}, K^{\prime}, V^{\prime}).
\end{equation}

\begin{figure*}[t]
  \centering    
  \includegraphics[width=\linewidth]{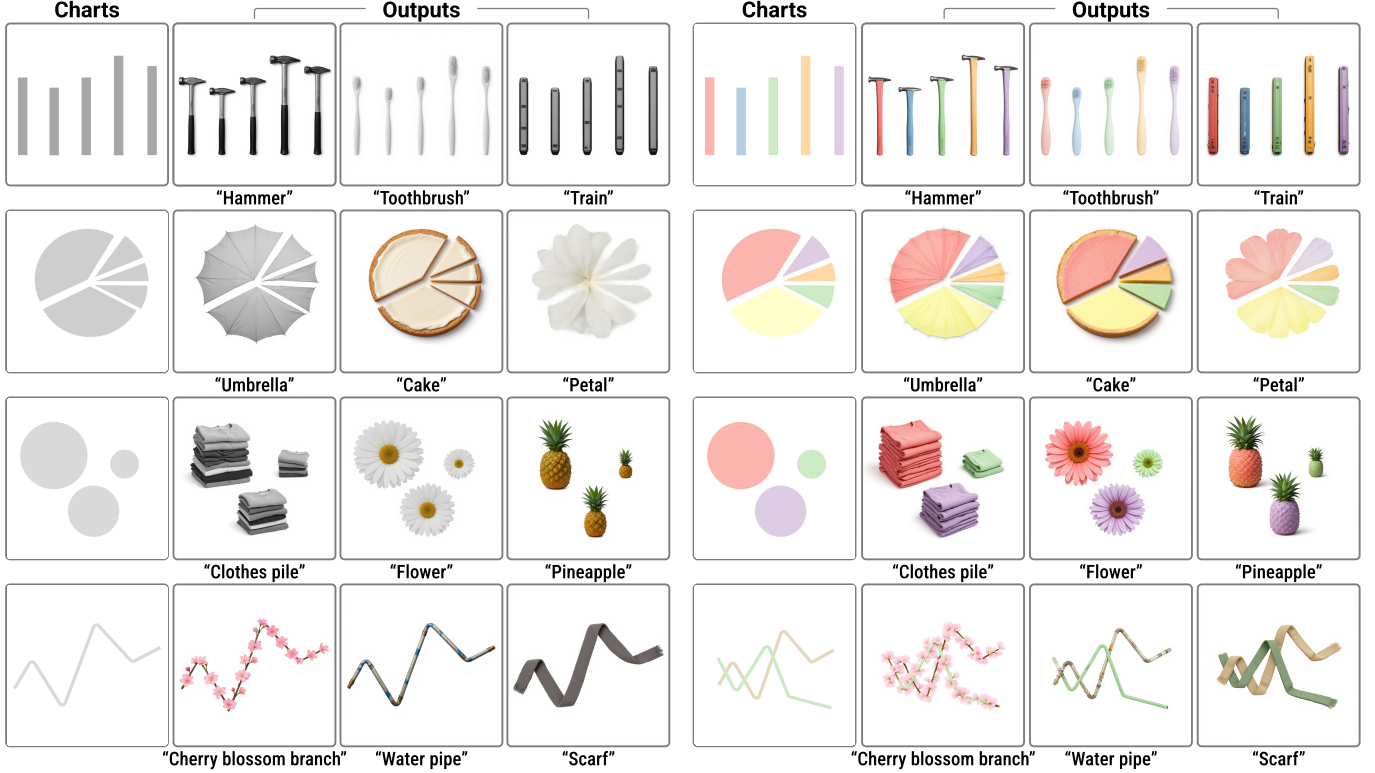}
    \caption{Qualitative Results. Our method generates diverse pictorial charts, preserving data-encoding colors and spatial structure during semantic synthesis.
    }
  \label{fig:results}
\end{figure*}

\section{Evaluation}

We validate our dual-conditioned generation framework through a comprehensive evaluation. This includes qualitative demonstrations of expressive semantic synthesis across diverse chart types, quantitative benchmarking against state-of-the-art controllable generation and DiT editing baselines to verify strong structural fidelity, and component-wise ablation studies to validate our architectural design choices.

\subsection{Qualitative Results}

\begin{figure}
    \centering    
  \includegraphics[width=\linewidth]{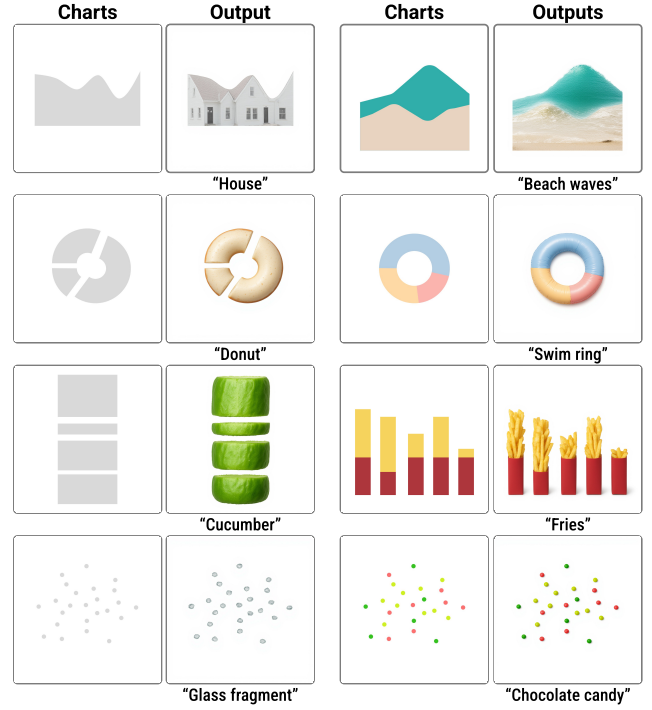}
    \caption{
    Generalization Across Chart Topologies.
    Examples of other chart types include area charts, donut charts, stacked bar charts, and scatter plots.
    }
  \label{fig:more example}
\end{figure}

\paragraph{Generative Performance}
Fig.~\ref{fig:results} and Fig.~\ref{fig:more example} demonstrate the versatility of our method across diverse chart types and multi-color encoding schemes. 
Our method exhibits strong structural fidelity to the input data, accurately preserving both the spatial structure and categorical color mappings, while faithfully synthesizing the target semantic visual referents defined by the prompt.

\begin{figure}
    \centering    
  \includegraphics[width=\linewidth]{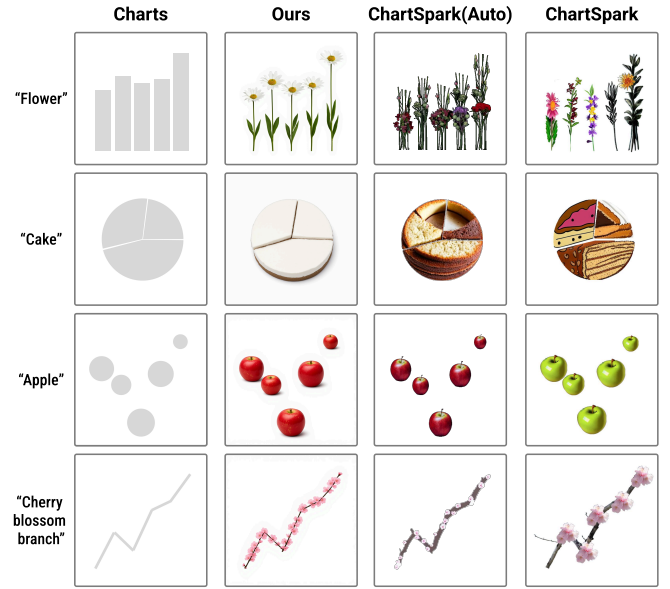}
    \caption{
    Comparison with Domain-Specific Baselines. 
    Qualitative evaluation against both the autonomous (third column) and user-interactive (fourth column) modes of ChartSpark~\cite{10.1109/TVCG.2023.3326913}.
    Compared to both modes, our method achieves superior structural fidelity and more cohesive semantic synthesis without requiring manual intervention.
    }
  \label{fig:additional-comparison}
\end{figure}

\begin{figure*}[t]
  \centering    
  \includegraphics[width=\linewidth]{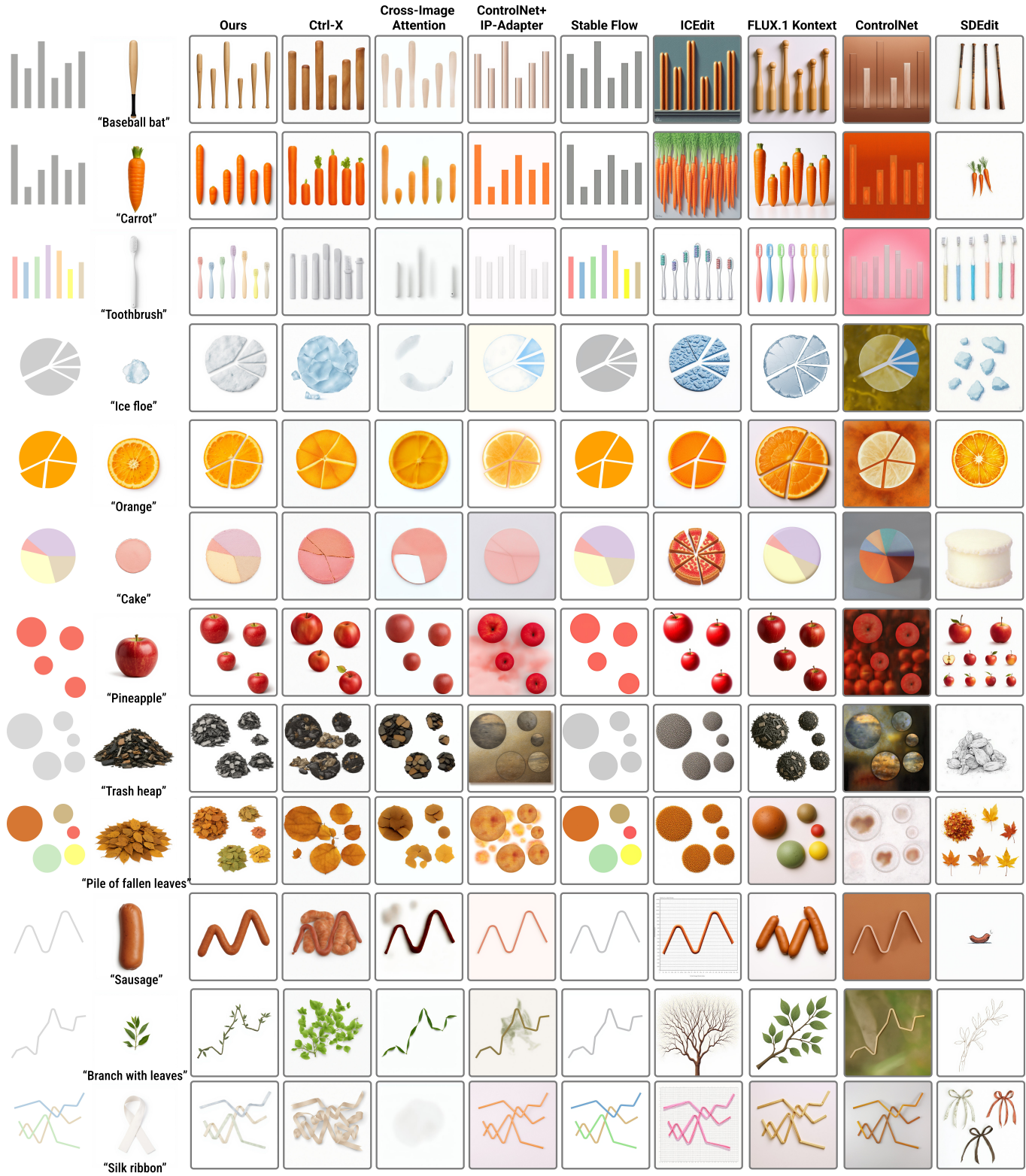}
    \caption{Qualitative Comparison. We compare our method against eight baselines for controllable generation and image editing. 
    The first three columns feature reference-guided methods utilizing an additional semantic reference image, while the remaining columns display text-guided methods relying solely on the source chart and prompt. Across these diverse conditions, our method achieves the optimal balance, performing expressive semantic synthesis while strictly preserving the structural fidelity of the original chart.
    }
  \label{fig:compare}
\end{figure*}

\paragraph{Baseline Comparisons}
In Fig.~\ref{fig:compare}, we benchmark our approach against a spectrum of methods that prioritize structural consistency during semantic transformation. 
We classify these baselines into two categories:
(1) \textit{Controllable Generation}, which includes training-based frameworks (ControlNet, IP-Adapter~\cite{IP-Adapter}), training-free attention-modulation methods (Ctrl-X, CIA), and domain-specific pictorial generation systems (ChartSpark~\cite{10.1109/TVCG.2023.3326913}, where Fig.~\ref{fig:additional-comparison} evaluates both its fully automated and interactively refined outputs).
(2) \textit{Image Editing}, which encompasses noise-based editing (SDEdit), inversion-based methods (Stable Flow), and context-aware DiT approaches (ICEdit, FLUX.1 Kontext).
As evidenced by the visual comparisons, our method achieves a superior balance between structural constraints and expressive generation. 
While baselines frequently suffer from either structural drift (compromising the data encoding) or semantic rigidity (applying superficial textures rather than synthesizing organic objects), our approach yields visual referents that are cohesive and natural, yet quantitatively faithful to the source chart.

\subsection{Quantitative Results}

\paragraph{Benchmark Construction}
To evaluate our method, we established a bespoke benchmark comprising 160 distinct chart–object pairs, as no existing dataset addresses this specific task. 
The benchmark spans four primary chart types (Bar, Line, Pie, Bubble), each paired with 10 distinct semantic objects. 
For every object, we generated four random chart instances to ensure structural diversity. 
To facilitate fair comparison with controllable generation baselines, we also synthesized corresponding semantic reference images. 
All prompts were curated via a hybrid pipeline of Large Language Models (LLMs) and manual verification, categorized into target prompts (for generation) and editing prompts (for modification) to cover diverse usage scenarios.

\begin{table}[b]
    \centering
    \setlength{\tabcolsep}{2.5pt}
    \caption{Quantitative Comparison. Evaluation of structural (SSIM, PSNR, DINO), perceptual (LPIPS), and semantic (CLIP) metrics. Our method demonstrates the optimal trade-off, minimizing structural deviation while maximizing expressive semantic synthesis.}
    \small
    \begin{tabular}{lrrrrr}
    \toprule
        \multirow{2}{*}{Methods} 
        & \multicolumn{3}{c}{Structure} 
        & \multicolumn{1}{c}{Perceptual} 
        & \multicolumn{1}{c}{Semantic} \\
        \cmidrule(lr){2-4} \cmidrule(lr){5-5} \cmidrule(lr){6-6}
         & SSIM~$\uparrow$ 
         & PSNR~$\uparrow$ 
         & DINO~$\downarrow$ 
         & LPIPS~$\downarrow$ 
         & CLIP~$\uparrow$ \\ 
    \midrule
        Stable Flow & \textbf{.99} & \textbf{32.91} & \textbf{.02} & \textbf{.02} & .22 \\ 
        ControlNet+IP-Adapter & \underline{.88} & 17.00 & .13 & .29 & .23 \\
        ControlNet  & .67 & 8.56 & .20 & .68 & .23 \\ 
        CIA & .90 & 19.69 & .13 & .23 & .24 \\ 
        ChartSpark (Auto) & .71 & 15.82 & .14 & .38 & \underline{.25} \\
        Ctrl-X & .78 & 15.17 & .21 & .38 & \textbf{.26} \\
        ICEdit & .51 & 9.24 & .30 & .78 & \textbf{.26} \\ 
        FLUX.1 Kontext & .77 & 14.85 & .16 & .37 & \textbf{.26} \\ 
        SDEdit & .84 & 17.03 & .23 & .37 & \textbf{.26} \\ 
    \midrule
        Ours & .86 & \underline{20.47} & \underline{.08} & \underline{.22} & \textbf{.26} \\
    \bottomrule
    \end{tabular}
    \label{tab:quantitative}
\end{table}

\paragraph{Evaluation Metrics}
We assess performance across two orthogonal axes: structural fidelity and semantic quality. 
To quantify structural preservation, we employ SSIM~\cite{SSIM}, PSNR, and the DINO self-similarity distance~\cite{Ctrl-X}, which measures the integrity of deep visual features. 
For semantic transformation, we utilize LPIPS~\cite{LPIPS} to measure perceptual deviation, indicating the synthesis quality and deviation from the abstract geometric primitives.
Furthermore, to ensure robust semantic evaluation unaffected by prompt phrasing, we compute CLIP Similarity~\cite{CLIP} directly between the generated image and the canonical object class label.

\paragraph{Comparative Analysis}
Table~\ref{tab:quantitative} details the comprehensive quantitative results. 
In terms of structural preservation, our method outperforms most baselines, second only to ControlNet (which imposes rigid, often unnatural constraints) and Stable Flow (which exhibits negligible editability). 
In terms of semantic synthesis, our approach achieves parity with methods that prioritize pure generation. 
Consequently, our framework demonstrates the optimal balance, securing strong structural fidelity without compromising the semantic richness required for expressive pictorial charts.

\begin{figure}[t]
  \centering    
  \includegraphics[width=1.\linewidth]{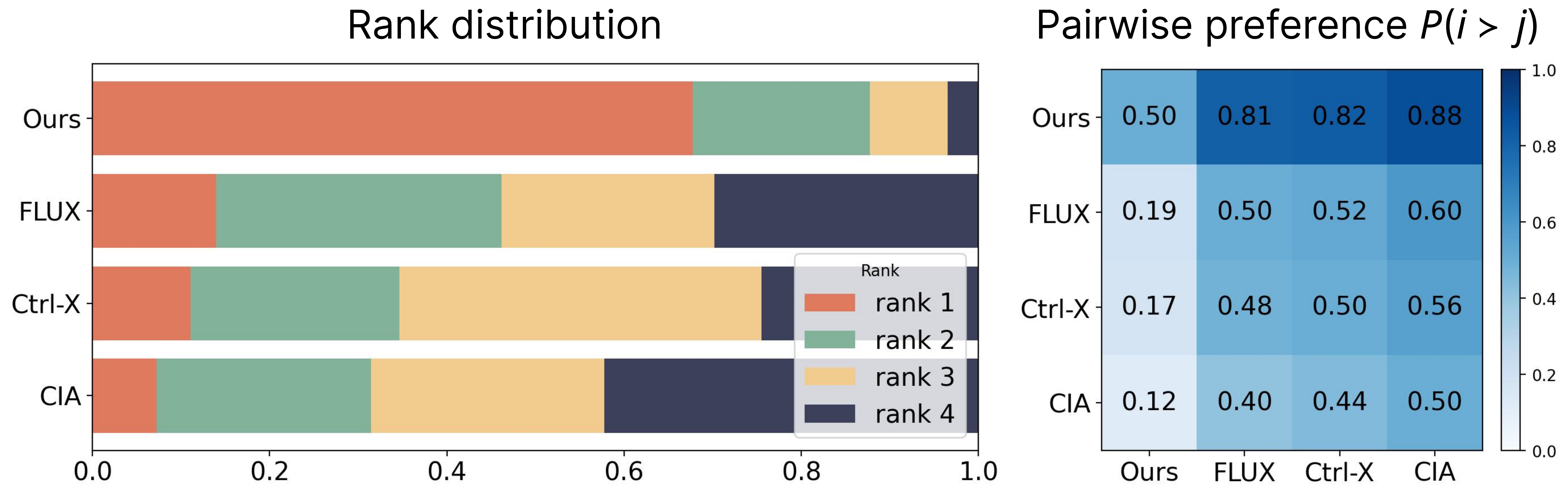}
    \caption{Visualization of User Study Results.
    Rank distribution (left) and pairwise preference heat map (right) comparing our method against Ctrl-X, CIA, and FLUX.1 Kontext. The results demonstrate a significant user preference for our dual-conditioned method in balancing structural fidelity with high-quality semantic synthesis.
    }
  \label{fig:userstudy}
\end{figure}

\paragraph{User Study}
To validate perceptual quality, we conducted a user study with 56 participants. 
We selected 20 diverse input charts and generated results using four methods: Ours, CIA, Ctrl-X, and FLUX.1 Kontext, yielding 80 total stimuli.
These baselines were chosen because they represent attention-based controllable generation (CIA, Ctrl-X), similar to our inference logic, while FLUX.1 Kontext serves as our backbone architecture.
Participants ranked the randomized outputs based on two criteria: (1) consistency with the original chart (spatial structure and data-encoding colors), and (2) effectiveness of the semantic transformation. 
We employed a rigorous attention check to ensure data quality, which all 56 participants passed. 
Given the ordinal nature of the ranking data, we analyzed the results using a Friedman test, revealing a statistically significant difference ($\chi^2(3) = 974.450, p < .001$). Post-hoc Wilcoxon signed-rank tests with Bonferroni correction ($\alpha \approx .017$) confirmed that our method significantly outperforms all baselines: CIA ($Z = -23.622, p < .001$), Ctrl-X ($Z = -21.949, p < .001$), and FLUX.1 Kontext ($Z = -21.015, p < .001$). Participants' voting demonstrates a strong consensus, with our method ranked as the top outcome in 67.8\% of trials (see Fig.~\ref{fig:userstudy} and Table~\ref{tab:userstudy}).

\begin{table}[t]
    \centering
    \caption{User Preference Statistics. Participant rankings demonstrate a strong consensus, with our method preferred (67.8\% of trials) over baselines.
    }
    \small
    \begin{tabular}{lrrrrr}
    \toprule
        Methods & 1st & 2nd & 3rd & 4th & Top-1 Rate \\ 
    \toprule
        CIA & 81 & 271 & 295 & 473 & 7.2\% \\
        Ctrl-X & 124 & 264 & 458 & 274 & 11.1\% \\
        FLUX.1 Kontext & 156 & 361 & 269 & 334 & 13.9\% \\ 
    \midrule
        Ours & 759 & 224 & 98 & 39 & \textbf{67.8\%} \\
    \bottomrule
    \end{tabular}
    \label{tab:userstudy}
\end{table}

\begin{figure}[b]
  \centering    
  \includegraphics[width=1.\linewidth]{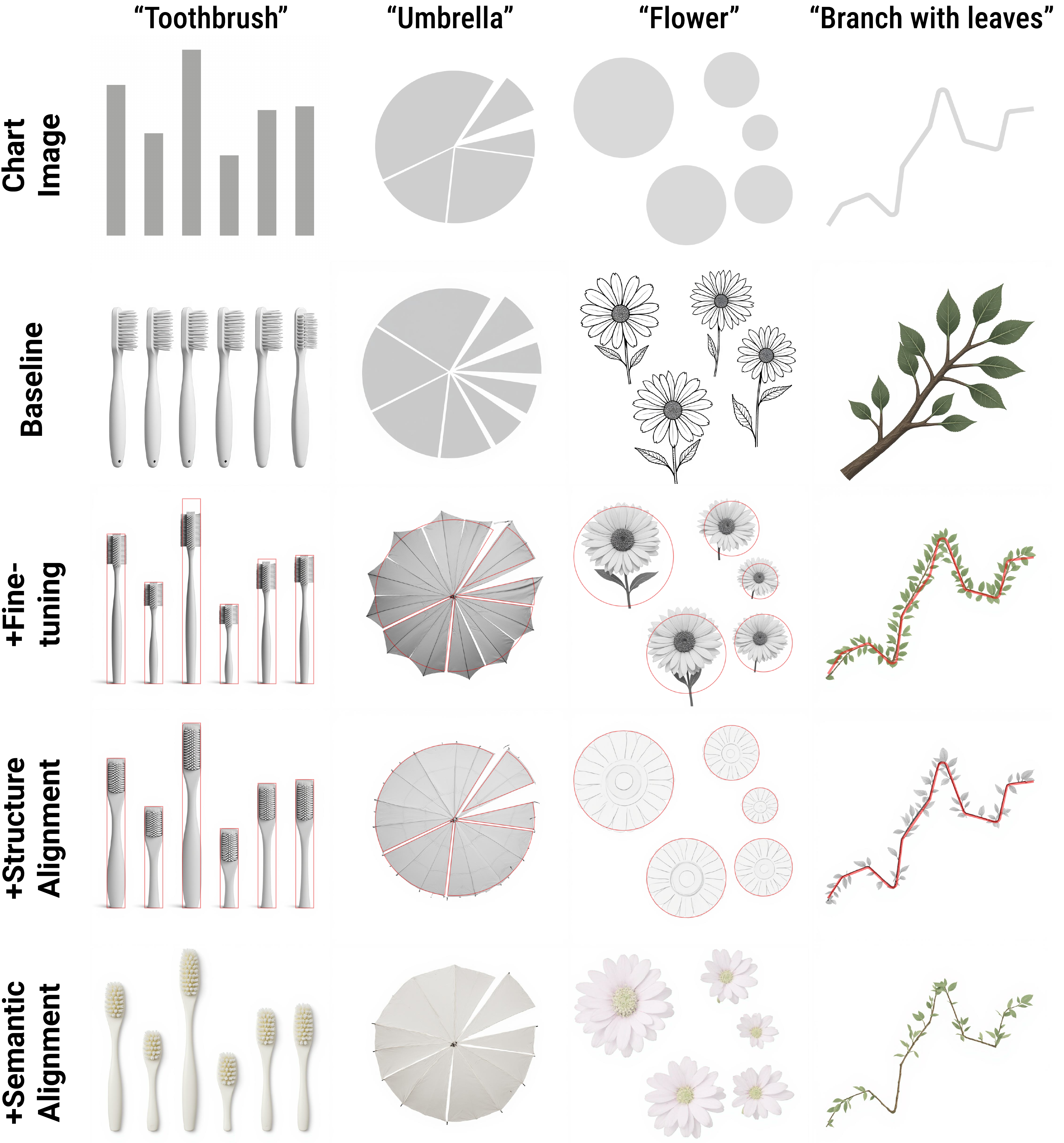}
    \caption{Ablation Study. 
    Columns demonstrate the progressive integration of specific modules across different themes (rows).
    The full method (bottom row) ensures both strict structural alignment and robust semantic synthesis.
    }
  \label{fig:ablation}
\end{figure}

\subsection{Ablation Studies}

We perform a component-wise ablation to validate the necessity of each stage in our method. 
The comparisons are presented in Fig.~\ref{fig:ablation}.

\paragraph{Impact of LoRA Fine-tuning}
The base FLUX.1 Kontext model lacks specific domain knowledge of statistical graphics. 
Without LoRA fine-tuning, the model fails to interpret abstract chart primitives, resulting in significant structural misalignment and generic, low-fidelity visual referents. 
Fine-tuning bridges this domain gap, enabling the model to correctly perceive global chart topology and generate expressive semantic associations.

\paragraph{Necessity of Structure Alignment} 
While fine-tuning establishes a rough global layout, it is insufficient to preserve the strong structural fidelity required for statistical charts. 
As seen in Fig.~\ref{fig:ablation}, relying solely on the learned prior often leads to local structural deviations that corrupt the data encoding. 
The integration of structural control is therefore critical as it enforces strict fidelity by anchoring the generated results directly to the spatial structure of the source chart, correcting the deviations.

\paragraph{Necessity of Semantic Alignment}
The explicit injection of structural constraints can sometimes degrade the organic quality of the output, introducing chart-like artifacts or suppressing texture details. 
Semantic control compensates for this by re-injecting high-fidelity semantic features from the reference image. 
This step is essential for resolving artifacts and ensuring the final output achieves the expressive semantic synthesis intended by the user.

\paragraph{Role of DIFT in Feature Transfer}
Finally, we validate the use of diffusion features for correspondence. 
As illustrated in Fig.~\ref{fig:appdift}, relying on naive attention matching (without DIFT) results in erroneous feature registration. This leads to structural incoherence, characterized by ghosting artifacts and severe spatial misalignment. 
Conversely, DIFT-based correspondence ensures precise feature matching, enabling robust semantic synthesis without visual degradation.

\section{Conclusions, Limitations and Future Work}

We presented a generative framework for pictorial chart generation that reframes the task as a problem of semantic-structural alignment, rather than superficial style transfer or decorative enhancement. 
Instead of treating statistical graphics as unconstrained domains for style transfer, our approach explicitly partitions the generative process into structural invariants and semantic variables.
We distinguish between the spatial structure, which encodes the underlying data and must be strictly preserved, and the semantics, which govern the expressive visual referents of the chart elements. This separation is operationalized through two complementary diffusion-feature-based mechanisms: Structural DIFT and Semantic DIFT.
Together, they independently regulate structural fidelity and semantic synthesis within a unified generative process. In doing so, this work advances diffusion features from a passive tool for correspondence analysis into an active control mechanism.
This enables fine-grained accountability during generation, allowing expressive semantic variation without compromising quantitative chart integrity.

\begin{figure}[t]
  \centering    
  \includegraphics[width=\linewidth]{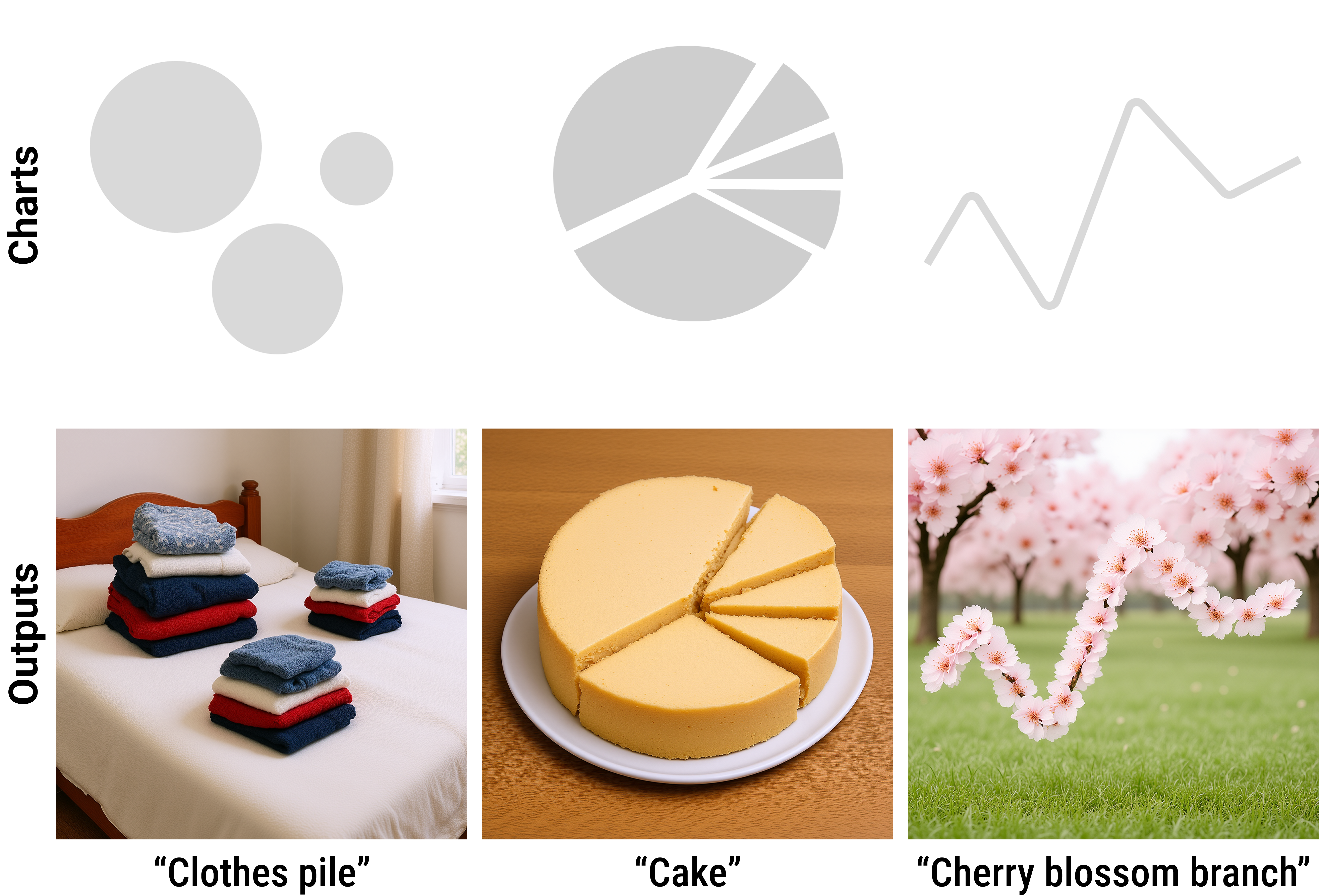}
    \caption{Future Explorations: Holistic Scene Generation. 
    Contextually coherent backgrounds are currently achieved by expanding the base text prompt. 
    Because explicit background control is not an objective of the current method, holistic scene generation relies heavily on the underlying model's prior, which can be disrupted by the chart-focused alignment mechanism.
    These preliminary results demonstrate the potential for full-scene data storytelling, underscoring the need for unified control mechanisms that govern background synthesis without compromising analytical legibility.
    }
  \label{fig:background}
\end{figure}

Although we have demonstrated the effectiveness of our dual-conditioned method, several limitations remain. 
First, enforcing strict structural alignment can occasionally bias generation toward chart-derived features, which may interfere with the intended semantic synthesis and disrupt contextual background generation. 
Consequently, achieving optimal semantic fidelity and background quality still partially depends on the expressive capacity of the fine-tuned base model and prompt specification. 
Second, since semantic alignment relies heavily on the reference image, synthesis quality may degrade when the reference's viewpoint or spatial layout differs substantially from the target chart primitives.

Currently, our approach relies on transformation capabilities introduced through model fine-tuning, inheriting the limitations associated with training-based adaptation. An important direction for future work is the development of zero-shot, training-free generation methods.
Enabling the semantic transformation of abstract geometric charts purely through inference-time attention control would further strengthen the separation between structural fidelity and semantic expressiveness.
Finally, we aim to extend the current framework from isolated chart-element transformation to holistic scene generation. This includes explicit and controllable background synthesis that remains contextually compatible with the chart's semantics without weakening data readability, as illustrated in Fig.~\ref{fig:background}.
In conclusion, by proving that generative priors can be rigorously constrained through diffusion-feature alignment, this work provides a stepping stone toward the next generation of expressive, data-driven visual storytelling.

\begin{acks}

This work was supported in part by ICFCRT (W2441020), Guangdong Basic and Applied Basic Research Foundation (2023B1515120026), Guangdong Natural Science Foundation (2026A1515010423), Shenzhen Science and Technology Program (KQTD20210811090044003, KJZD20240903100022028), NSFC (62472288), IITP Grant funded by MSIT (RS-2020-II201361), and Scientific Development Funds from Shenzhen University.

\end{acks}

\bibliographystyle{ACM-Reference-Format}
\bibliography{main}

\newpage

\newpage

\end{document}